\documentclass[aps,prl,twocolumn,superscriptaddress]{revtex4-2}
\usepackage{amsmath,amssymb,bm,physics}
\usepackage{graphicx}
\usepackage{hyperref}
\usepackage{color}

\begin{document}

\title{Exact Nagaoka-to-spiral transition in the doped infinite-$U$ triangular lattice}

\author{Yang Zhang}
\affiliation{Department of Physics and Astronomy, University of Tennessee, Knoxville, TN 37996, USA}
\affiliation{Min H.\ Kao Department of Electrical Engineering and Computer Science, University of Tennessee, Knoxville, TN 37996, USA}

\author{Cristian D.\ Batista}
\affiliation{Department of Physics and Astronomy, University of Tennessee, Knoxville, TN 37996, USA}
\affiliation{Shull Wollan Center, Oak Ridge National Laboratory, Oak Ridge, TN 37831, USA}

\begin{abstract}

We study the single-hole infinite-$U$ Hubbard model on the triangular lattice with nearest- and next-nearest-neighbor hoppings $t_1$ and $t_2$. A frustrating $t_2$ destabilizes Nagaoka ferromagnetism through a Lifshitz transition to a long-wavelength coplanar spiral at $t_{2,c}/t_1=-0.182$. We determine the critical point analytically by reducing the many-body problem to an effective two-body hole--magnon scattering problem in which the hard-core hole--magnon constraint is treated exactly. Numerical calculations confirm the spiral ground state and reveal a coherent spin polaron with quasiparticle weight $Z\simeq0.92$ at the Lifshitz-shifted momentum $\mathbf{K}=-\mathbf{Q}^*$. These results establish the first quantum instability of the triangular-lattice Nagaoka ferromagnet and provide the magnetic baseline for magnon-mediated pairing at finite doping.
\end{abstract}

\maketitle

\textbf{Introduction.} Kinetic frustration, whereby hole motion on geometrically
frustrated lattices generates effective antiferromagnetic
interactions from purely repulsive Coulomb
forces~\cite{Haerter2005,Sposetti2014,Isaev2010},
plays a central role in doped Mott insulators.
On the triangular lattice, it competes directly with
Nagaoka ferromagnetism~\cite{Nagaoka1966,Thouless1965},
giving rise to spin
polarons~\cite{SSZhang2018,Davydova2023},
resonating-valence-bond spin
liquids~\cite{Glittum2025},
and magnon-mediated
superconductivity~\cite{SSZhang2018,morera2024attraction,Nazaryan2024,Zhu2026,YZhang2025}.
Recent observations of kinetic magnetism in moir\'e
materials~\cite{Ciorciaro2023,Tao2024} and cold-atom quantum
simulators~\cite{Xu2023,Lebrat2024,Prichard2024}
have renewed interest in these frustration-driven magnetic
phases.

A central open question is how Nagaoka ferromagnetism evolves as kinetic
frustration is introduced. Although the Nagaoka theorem rigorously
establishes a fully polarized ground state for a single hole with
nearest-neighbor (NN) hopping
$t_1>0$~\cite{Nagaoka1966,Tasaki1998,Hanisch1997}, its fate in the
presence of frustrating next-nearest-neighbor (NNN) hopping,
$t_2<0$, has been explored primarily through DMRG and semiclassical
studies of square-lattice
variants~\cite{Lisandrini2017,Sharma2024,PereiraMueller2025}. The exact
critical hopping $t_{2,c}$ and the order of the transition remain
unknown.
 A semiclassical analysis reveals that the fully polarized
state minimizes the hole kinetic energy for any $t_1$ and $t_2$,
implying that any instability must involve quantum fluctuations.
The resulting transition has the character of a Lifshitz
instability in which the minimum of the magnon dispersion shifts
continuously from $\mathbf Q=0$ to a finite wave vector.

In this work, we identify the quantum instability of the
triangular-lattice Nagaoka ferromagnet as a long-wavelength spin
spiral and determine the exact critical value
$t_{2,c}=-0.182\,t_1$. By reducing the many-body problem to an
exactly solvable two-body hole--magnon scattering problem that
treats the hard-core hole--magnon constraint exactly, we obtain an
analytical determination of the transition point, independently
confirmed by ED and flux-insertion studies. Within the spiral phase,
a variational calculation reveals a coherent spin polaron with
quasiparticle weight $Z\simeq0.92$ at the Lifshitz-shifted momentum
$\mathbf{K}=-\mathbf{Q}^*$.

The spiral phase has implications beyond the single-hole limit. As
$\mathbf{Q}\to0$, the saturation field becomes arbitrarily small, while
below saturation holes are known to bind tightly by forming three-body
states with a single
magnon~\cite{SSZhang2018,morera2024attraction,Nazaryan2024,Zhu2026}. The spiral wavevector
$\mathbf{Q}$ selects the pairing symmetry through the nesting geometry,
whereas the spin stiffness sets the effective attraction. Our
single-hole results thus provide the quantitative magnetic baseline for
finite-doping theories.

\textbf{Model and spiral energy.}
We consider a single hole in the $U=\infty$ Hubbard model on the
triangular lattice with NN hopping $t_1>0$ and NNN hopping $t_2$
(electron-doped convention; $t_2<0$ frustrates the ferromagnet).
The bare hole dispersion in the ferromagnetic (FM) background is
\begin{equation}
\epsilon_{\bm k}
=
-\sum_{\bm\delta} t_{\bm\delta} e^{i\bm k\cdot\bm\delta},
\label{eq:epsilon}
\end{equation}
where
$\bm\delta=\{\pm\mathbf a_{1,2,3},\pm\mathbf b_{1,2,3}\}$ are the NN
and NNN lattice vectors. The FM ground state corresponds to
$\mathbf{k}=\Gamma$, with energy
$
E_{\rm FM}=-6t_1-6t_2.
$

To describe the spiral state, we transform to the local frame of a
coplanar spiral with wavevector $\bm Q$. The site-dependent rotation
generates spin-conserving and spin-flip hoppings proportional to
$\cos(\bm Q\!\cdot\!\bm\delta/2)$ and
$\sin(\bm Q\!\cdot\!\bm\delta/2)$, respectively. In terms of spinless
hole ($h_i$) and magnon ($a_i$) operators, the one-hole Hamiltonian
becomes
\begin{equation}
H(\bm Q)=H_h(\bm Q)+H_{h\text{-}m}^{\rm hyb}(\bm Q)
+H_{h\text{-}m}^{\rm int}(\bm Q).
\end{equation}
In the classical ($\bm Q\to0$, equivalently $S\to\infty$) limit, the
leading term describes hole hopping in the spiral background,
\begin{equation}
H_h(\bm Q)=
-\sum_{j,\bm\delta}t_{\bm\delta}
\cos\!\left(\frac{\bm Q\!\cdot\!\bm\delta}{2}\right)
h_{{\bm R}_j+\bm\delta}^\dagger h_{{\bm R}_j},
\end{equation}
with bare dispersion
\begin{equation}
\varepsilon_0(\bm k;\bm Q)=
-\sum_{\bm\delta}t_{\bm\delta}
\cos\!\left(\frac{\bm Q\!\cdot\!\bm\delta}{2}\right)
e^{i\bm k\cdot\bm\delta}.
\end{equation}

The spin-flip term, $H_{h\text{-}m}^{\rm hyb}(\bm Q)$, hybridizes the
single-hole and one-magnon sectors:
\begin{align}
H_{h\text{-}m}^{\rm hyb}
&=
- i \sum_{j,\bm\delta} t_{\bm\delta}
\sin\!\left(\frac{\bm Q\cdot\bm\delta}{2}\right)
\, h_{{\bm R}_j+\bm\delta}^\dagger h_{{\bm R}_j}
\left(a_{{\bm R}_j+\bm\delta} - a_{{\bm R}_j}^\dagger\right) \nonumber \\
&=
\frac{1}{\sqrt N}
\sum_{\bm k,\bm q}
\Big[
M_{\bm q}(\bm k,\bm Q)\,
h_{\bm k-\bm q}^\dagger h_{\bm k}\, a_{\bm q}^\dagger
+\text{H.c.}
\Big],
\end{align}
with
\begin{equation}
M_{\bm q}(\bm k,\bm Q)
=
i \sum_{\bm\delta} t_{\bm\delta}
\sin\!\left(\frac{\bm Q\cdot\bm\delta}{2}\right)
e^{i\bm k\cdot\bm\delta}
\left(1-e^{-i\bm q\cdot\bm\delta}\right).
\end{equation}
The vertex satisfies
$
M_{\bm q}(\bm k,\bm Q)=\mathcal{O}(Q)
$
as $Q\equiv|\bm Q|\to0$, providing a controlled small-$Q$
expansion of the ferromagnetic-to-spiral transition. The regular
hole--magnon interaction is therefore perturbative, while states with
two or more magnons contribute only at higher orders and do not affect
the energy at order $Q^2$.

The hard-core hole--magnon constraint is incorporated exactly by
introducing an on-site repulsion $V$ and taking the limit
$V\to\infty$:
\begin{eqnarray}
H_{h\text{-}m}^{\rm int}(\bm Q)
&=&-
\sum_{j,\bm\delta}
t_{\bm\delta}\cos\!\left(\frac{\bm Q\cdot\bm\delta}{2}\right)
h_{{\bm R}_j}^\dagger 
a_{{\bm R}_j+\bm\delta}^\dagger h_{{\bm R}_j+\bm\delta} a_{{\bm R}_j},
\nonumber \\
&+& V \sum_j h_{{\bm R}_j}^\dagger h_{{\bm R}_j} a_{{\bm R}_j}^\dagger  a_{{\bm R}_j}
\qquad V\to\infty.
\end{eqnarray}
The small-$Q$ expansion thus reduces the many-body problem to a
two-body hole--magnon scattering problem with a hard-core constraint.

The classical spiral energy is obtained by minimizing
$\varepsilon_0(\bm k;\bm Q)$ with respect to $\bm k$ for each
$\bm Q$. Since the hole--magnon vertex satisfies
$M_{\bm q}(\bm k,\bm Q)= \mathcal{O}(Q)$,
the leading quantum correction arises entirely from virtual
one-magnon processes. The energy is minimized for
$\bm Q$ along a $\Gamma$--$M$ direction and admits the expansion
\begin{equation}\label{eq:EQ}
E(Q)-E_{\rm FM}
=
(\alpha_{\rm cl}-\Pi_0)Q^2
+c_4Q^4
+  \mathcal{O} (Q^6),
\end{equation}
where
$\alpha_{\rm cl}=(3t_1+9t_2)/8$ is the classical spin stiffness and
$\Pi_0Q^2$ is the order-$Q^2$ correction from virtual one-magnon
processes. The total quadratic coefficient,
$c_2\equiv\alpha_{\rm cl}-\Pi_0$, vanishes at the transition.
Higher-magnon sectors contribute only at higher orders in
$Q$ and therefore do not renormalize the spiral stiffness.

Since only the one-magnon sector contributes at $\mathcal{O}(Q^2)$, the
small-$Q$ expansion reduces to a two-body hole--magnon scattering
problem with a hard-core constraint. Identifying the zero-magnon state
with an auxiliary relative-coordinate state
$|\bm r=\bm 0\rangle$ maps the problem onto single-particle scattering
in the relative coordinate $\bm r$.

The  odd-parity bond states
\begin{eqnarray}
|u_1\rangle &=&
\sum_{\mu=1}^{3}
(\hat{\mathbf Q}\!\cdot\!\mathbf a_\mu)
\left(|\mathbf a_\mu\rangle-|-\mathbf a_\mu\rangle\right),
\nonumber\\
|u_2\rangle &=&
\sum_{\mu=1}^{3}
(\hat{\mathbf Q}\!\cdot\!\mathbf b_\mu)
\left(|\mathbf b_\mu\rangle-|-\mathbf b_\mu\rangle\right),
\end{eqnarray}
span the hybridization channel as $Q\to0$ (see SM~\cite{SM}).
Projection onto this subspace yields an exact $2\times2$
Lippmann--Schwinger equation [see Supplemental
Material (SM)~\cite{SM}] and the self-energy
\begin{equation}
\Sigma(E,Q)
=
\frac14 Q^2\,{\bm t}^{T}
{\cal G}(E)\,
{\bm t}
+\mathcal{O}(Q^4),
\end{equation}
where
\[
{\bm t}
=
\left(
\sqrt{\langle u_1|u_1\rangle}\,t_1,
\sqrt{\langle u_2|u_2\rangle}\,t_2
\right)^T
\]
is the vector of NN and NNN hopping amplitudes in the odd-parity bond
basis, and
\[
{\cal G}(E)
=
\bigl[{\cal G}^{0}(E)^{-1}-W\bigr]^{-1}
\]
is the corresponding projected two-body resolvent.
Substituting into Eq.~(\ref{eq:EQ}) yields 
\begin{equation} \Pi_0 = -\frac14\, {\bm t}^{T} {\cal G}(E_{\rm FM}) {\bm t}. \label{eq:Pi0final} 
\end{equation}
The explicit forms of ${\cal G}^0(E)$ and $W$, together
with the derivation of Eq.~(\ref{eq:Pi0final}), are given
in the SM~\cite{SM}.
Evaluating Eq.~(\ref{eq:Pi0final}) at the critical point gives
$\Pi_0=0.170$ and $t_{2,c}/t_1=-0.182$. Relative to one-loop theory,
the exact treatment of the hard-core $dd$ channel suppresses
$\Pi_0$ by $\sim24\%$, from
$\Pi_0^{(1)}=0.225$ to $0.170$, shifting the critical hopping from
$t_{2,c}^{(1)}/t_1=-0.133$ to $-0.182$ and extending the Nagaoka phase
by $37\%$ (see SM~\cite{SM}).

\begin{figure}[t]
\includegraphics[width=\columnwidth]{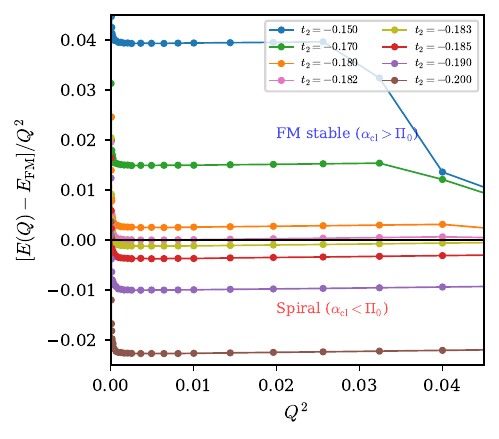}
\caption{Spiral stiffness $[E(Q) - E_\mathrm{FM}]/Q^2$ ($Q$ along
$\Gamma$--$M$) from the Trugman-01 method (0+1 magnon space) on
$100 \times 100$. The sign change between $t_2 = -0.182$ and
$-0.183$ pinpoints $t_{2,c}/t_1=-0.182$; the $Q\to 0$ plateau gives
$\alpha_\mathrm{cl} - \Pi_0$, while the bending at larger $Q$
reflects the positive one-magnon $\mathcal{O}(Q^4)$ stiffness.}\label{fig:EQ}
\end{figure}

As an independent check, we diagonalize the full $0{+}1$-magnon
Hamiltonian $H(\mathbf{Q})$ on a $100\times100$ lattice (the
Trugman-01 method). Because the $Q^2$ coefficient is determined
entirely by the one-magnon sector, this calculation yields the exact
spiral stiffness and a variational upper bound at finite $Q$.
Figure~\ref{fig:EQ} shows $[E(Q)-E_{\rm FM}]/Q^2$ for
$\mathbf{Q}$ along $\Gamma$--$M$. The sign change between
$t_2/t_1=-0.182$ and $-0.183$ pinpoints the critical point, in
excellent agreement with the analytical prediction. The $Q\to0$
plateau approaches $\alpha_{\rm cl}-\Pi_0$, while the upward
curvature reflects a positive one-magnon
$\mathcal{O}(Q^4)$ correction.

The transition order is determined by the quartic coefficient
$c_4$ in Eq.~\eqref{eq:EQ}. Since the spin-flip vertex satisfies
$M_{\bm q}=\mathcal{O}(Q)$, the $Q^4$ coefficient receives
contributions from at most two magnons,
\begin{equation}
c_4
=
-\beta_{\rm cl}
+c_4^{\rm 1\text{-}mag}
+c_4^{\rm 2\text{-}mag},
\label{eq:c4}
\end{equation}
where
$-\beta_{\rm cl}=-(3t_1+27t_2)/512>0$
is the classical contribution,
$c_4^{\rm 1\text{-}mag}$ is the one-magnon (polaronic) correction,
and
$c_4^{\rm 2\text{-}mag}$
is the two-magnon contribution arising from the zero-point energy of
the spiral. Because it first appears at
$\mathcal{O}(Q^4)$, it does not affect $t_{2,c}$.

The Trugman-012 diagonalization ($0{+}1{+}2$-magnon space) gives
\[
c_4^{\mathrm{2\text{-}mag}}
=
c_4^{(012)}-c_4^{(0+1)},
\]
evaluated on the same lattice. We obtain a sizable correction,
$c_4^{\mathrm{2\text{-}mag}}\approx-0.011$
(see SM~\cite{SM}), comparable to the semiclassical
spin-wave estimate. Although it substantially reduces the quartic
coefficient, the net value remains positive,
$c_4\approx+0.006$--$+0.008$, because
$-\beta_{\rm cl}+c_4^{\rm 1\text{-}mag}\approx+0.017$--$+0.019$.
Accordingly, the optimal spiral wavevector grows continuously from
zero at $t_{2,c}$ [Fig.~\ref{fig:polaron}(b)], with no discontinuous
jump or competing finite-$Q$ minimum,
\begin{equation}
Q^*
=
\sqrt{\frac{\Pi_0-\alpha_{\rm cl}}{2c_4}}
\propto
\sqrt{|t_2-t_{2,c}|},
\label{eq:Qstar}
\end{equation}
confirming a continuous FM-to-spiral transition.

\textbf{Exact diagonalization.}
We confirm the spiral instability using exact diagonalization (ED) on a
$3\sqrt{3}\times3\sqrt{3}$ torus ($N=27$). Sweeping all $S_z$ sectors on
six clusters ($N=20$--$30$) reveals an extended non-FM regime: for
$t_2<-0.25$, the ground state has minimal spin ($0$ or $1/2$) on all
clusters with full NNN coordination ($L_x\ge4$), while the Nagaoka FM
becomes unstable at
$t_{2,c}^{\rm ED}\approx-0.20$, consistent with the analytical value
$t_{2,c}/t_1=-0.182$ within finite-size corrections~(see SM \cite{SM}).

A second signature is the ground-state degeneracy of the
$3\sqrt{3}\times3\sqrt{3}$ cluster. For $t_2\le-0.25$, the two lowest
$\Gamma$ states are degenerate within machine precision and form the
$C_3$ $E$ co-representation with eigenvalues $\omega$ and $\omega^*$,
consistent with the incommensurate spiral partners
$\pm\mathbf Q^*$ folding onto $\Gamma$~(see \cite{SM}). The absence of this
degeneracy on rectangular clusters lacking $C_3$ symmetry rules out a
topological quantum-spin-liquid origin.

\begin{figure}[t]
\includegraphics[width=\columnwidth]{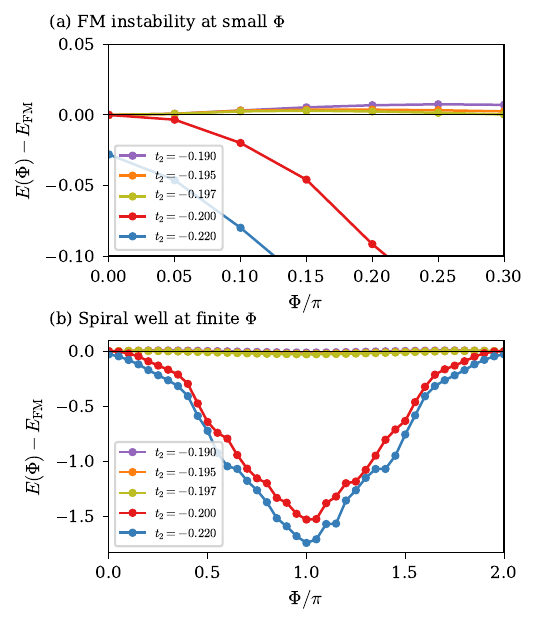}
\caption{Spectral flow on $3\sqrt{3}{\times}3\sqrt{3}$ ($N = 27$,
$S_z = 0$): the exact ground-state energy $E(\Phi) - E_\mathrm{FM}$
vs.\ flux $\Phi$ threading the torus, for $t_2/t_1 \in \{-0.190,
-0.195, -0.197, -0.200, -0.220\}$. (a)~Zoom near $\Phi = 0$:
$E(\Phi) - E_\mathrm{FM}$ stays $\geq 0$ for $t_2 \geq -0.197$ (FM
stable) and drops below zero at the smallest $\Phi$ for
$t_2 \leq -0.200$ (spiral instability), pinning the ED transition
between $-0.197$ and $-0.200$. (b)~Full $\Phi$ range: a spiral
energy well centered at $\Phi \approx \pi$ emerges for
$t_2 \leq -0.200$, with depth $\sim 1.5\,t_1$ at $t_2 = -0.200$
deepening to $\sim 1.7\,t_1$ at $t_2 = -0.220$, quantifying how far
the spiral ground state lies below
$E_\mathrm{FM}$.}\label{fig:ED}
\end{figure}

\begin{figure}[t]
\includegraphics[width=\columnwidth]{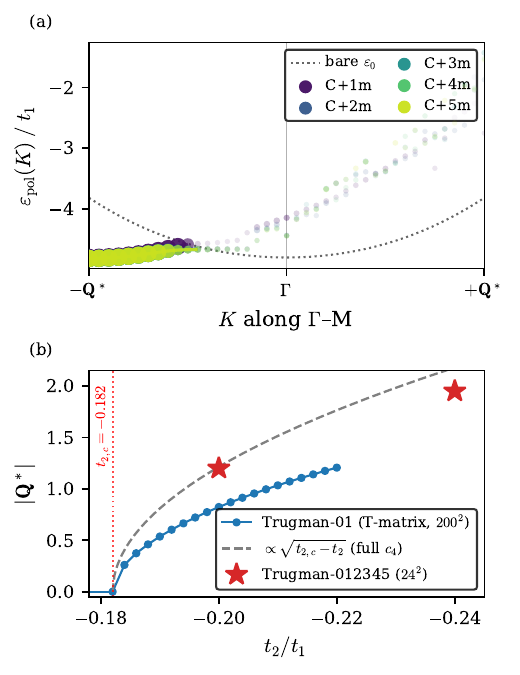}
\caption{(a)~Polaron dispersion in the spiral phase ($t_2/t_1=-0.20$) versus the
center-of-mass momentum $\mathbf{K}$ along the lab-frame path
$-\mathbf{Q}^*\!\to\!\Gamma\!\to\!+\mathbf{Q}^*$ ($|\mathbf{Q}^*|=1.2$),
for magnon truncations C+$1m,\ldots,$C+$5m$ (viridis). Marker size
$\propto Z(\mathbf{K})$; dotted: bare dispersion $\varepsilon_0$. A sharp
quasiparticle ($Z\simeq0.92$) at the Lifshitz-shifted momentum
$\mathbf{K}=-\mathbf{Q}^*$ evolves into a multi-magnon continuum beyond the
critical momentum $\mathbf{K}_c$, where the spiral becomes locally
unstable (see SM~\cite{SM}). The dispersion is already converged at the one-magnon
level, consistent with the suppression of higher-magnon processes for the
small spiral wavevector $Q^*$.
(b) Optimal magnitude of spiral wavevector, $Q^*$, versus $t_2/t_1$ along
$\Gamma$--$M$, obtained from
$Q^*=\arg\min_{\mathbf{Q}}E(\mathbf{Q})$. Blue:
one-magnon Trugman-01 $T$-matrix with $E(\mathbf{Q})$ computed by ED on
$200^2$ lattices ($24^2$ agrees to three digits). $Q^*$
vanishes in the FM phase and grows continuously from
$t_{2,c}=-0.182$ (red dotted), where $c_2=0$, confirming a second-order
transition. The gray dashed line is the analytical prediction,
Eq.~(\ref{eq:Qstar}), normalized at $t_2=-0.20$ to the full-quartic value
$\sqrt{-c_2/(2c_4)}=1.21$, with
$c_4=c_4^{(0+1)}+c_4^{2\text{-}mag}$. The red stars are the multi-magnon
Trugman-012345 variational minima over $(\mathbf{Q},\theta)$ ($24^2$,
$r=2$), yielding the $|\mathbf{Q}^*|=1.2$ used in (a).}\label{fig:polaron}
\end{figure}

Figures~\ref{fig:ED}(a),(b) show the spectral flow of the exact
$S_z=0$ ground-state energy under a flux $\Phi$ threaded through the
torus. Flux insertion twists the boundary conditions, probing spiral
states with arbitrarily small $Q$ near $\Gamma$. In
Fig.~\ref{fig:ED}(a), $E(\Phi)$ drops below $E_{\rm FM}$ at the
smallest nonzero flux for $t_2=-0.2$, identifying the first instability
of the FM toward a long-wavelength spiral~(see SM \cite{SM}). Over the full flux
range [Fig.~\ref{fig:ED}(b)], the spiral minimum near $\Phi\simeq\pi$
deepens from $\sim1.5\,t_1$ at $t_2=-0.200$ to $\sim1.7\,t_1$ at
$t_2=-0.220$, reflecting its increasing stability relative to the
ferromagnet.

\textbf{Coherent quasiparticle dispersion.}
Having established the spiral ground state, we next determine the nature
of its elementary quasiparticle. At $t_2/t_1=-0.20$, the variationally
optimized spiral wavevector lies along the $\Gamma$--$M$ direction with
magnitude $|\mathbf{Q}^*|=1.2$. The hole spectral function is
\begin{equation}
A(\mathbf K,\omega)
=
\sum_n Z_n(\mathbf K)\,
\delta\!\left(\omega-\omega_n(\mathbf K)\right),
\label{eq:Akw}
\end{equation}
where $\mathbf K$ is the hole momentum, and
$
Z_n(\mathbf K)
=
\left|
\langle n,\mathbf K
|c_{\mathbf K\sigma}|\Psi_0\rangle
\right|^2,
$
is computed by Lanczos diagonalization in a hybrid variational basis comprising
the Trugman-$01$ sector and radius-truncated multi-magnon states with up to
five magnons within distance $r=2$ of the hole (Hilbert space
$\sim10^6$)~\cite{SM}.

As expected for a small-$Q^*$ spiral, convergence with respect to the
magnon number is rapid near the band minimum, with all truncations
collapsing onto a single curve
($|{\rm C{+}5m}-{\rm C{+}3m}|<5\times10^{-3}t_1$).
Beyond a critical hole momentum $\mathbf K_c$, the quasiparticle weight
drops rapidly as the coherent pole merges into a multi-magnon
continuum, signaling that the $\mathbf Q^*$ spiral is no longer locally
stable for the propagating hole (see SM~\cite{SM}).

Remarkably, the spin polaron remains highly mobile despite its dressing,
with a bandwidth of $\sim9\,t_1$, only $25\%$ smaller than the bare value
($12\,t_1$). This weak renormalization contrasts sharply with the heavy
spin polarons of two-dimensional
antiferromagnets~\cite{BrinkmanRice1970,KaneLeeRead1989,MartinezHorsch1991}.
The origin is that $|\mathbf Q^*|\ll\pi/a$: the hole propagates through a
slowly twisting spin texture that remains locally ferromagnetic, rather
than a strongly fluctuating N\'eel background.

The Lifshitz shift leaves a clear spectroscopic fingerprint: the
quasiparticle minimum is displaced from $\Gamma$ to $-\mathbf Q^*$, with
Bragg replicas at $-\mathbf Q^*+n\mathbf Q^*$ in the finite-weight tails
of $A(\mathbf K,\omega)$. Together, the shifted band minimum and large
quasiparticle weight set the nesting geometry and interaction scale for
magnon-mediated pairing at finite doping.

\textbf{Conclusions.}
We have determined the exact phase boundary where the Nagaoka
ferromagnet of the doped infinite-$U$ triangular lattice transitions to
a long-wavelength spiral,
$t_{2,c}/t_1=-0.182$, and shown that the spiral phase hosts a coherent
spin polaron at the Lifshitz-shifted momentum $-\mathbf Q^*$. 
These results establish the magnetic
baseline for the doped triangular Hubbard model and provide the
foundation for magnon-mediated pairing at finite doping, where holes are
known to bind tightly to a single
magnon~\cite{SSZhang2018,Zhu2026,YZhang2025}. Natural extensions include
finite-$J$, finite-density, and Lifshitz-shifted Fermi-surface effects.

\textbf{Acknowledgments.} We thank Hantian Zhu and Shang-Shun Zhang for useful
discussions.  Computations were performed on the ISAAC cluster at
the University of Tennessee. C.D.B. acknowledges support from the U.S. Department of Energy, Office of Science, Office of Basic Energy Sciences, under Award Number DE-SC0022311.

\bibliography{refs}

\clearpage
\setcounter{equation}{0}
\setcounter{figure}{0}
\setcounter{table}{0}
\setcounter{section}{0}
\setcounter{secnumdepth}{3}
\renewcommand{\theequation}{S\arabic{equation}}
\renewcommand{\thefigure}{S\arabic{figure}}
\renewcommand{\thetable}{S\arabic{table}}
\renewcommand{\thesection}{S\Roman{section}}

\onecolumngrid
\begin{center}
{\large\textbf{Supplemental Material for ``Exact Nagaoka-to-spiral transition\\ in the doped infinite-$U$ triangular lattice''}}
\end{center}
\vspace{0.5em}
\twocolumngrid

\section{Model and lattice conventions}\label{sm:model}

We consider the $U=\infty$ Hubbard model~\cite{Nagaoka1966,Tasaki1998}
on the triangular lattice with one hole, corresponding to
$N_{\mathrm{el}}=N-1$ electrons on $N$ sites. At $J=0$, the Hamiltonian
is purely kinetic:
\begin{equation}
H
=
-\sum_{j,\boldsymbol{\delta},\sigma}
t_{\boldsymbol{\delta}}\,
\mathcal{P}\,
c_{\mathbf R_j+\boldsymbol{\delta},\sigma}^{\dagger}
c_{\mathbf R_j,\sigma}\,
\mathcal{P},
\label{eq:H}
\end{equation}
where $\mathcal{P}$ projects out states with double occupancy.
The NN lattice vectors are
\begin{equation}
\mathbf a_1=(1,0),\qquad
\mathbf a_2=\left(\frac{1}{2},\frac{\sqrt{3}}{2}\right),\qquad
\mathbf a_3=\left(-\frac{1}{2},\frac{\sqrt{3}}{2}\right),
\end{equation}
and the NNN lattice vectors are
\begin{equation}
\mathbf b_1=\mathbf a_1+\mathbf a_2,\qquad
\mathbf b_2=\mathbf a_2+\mathbf a_3,\qquad
\mathbf b_3=\mathbf a_3-\mathbf a_1.
\end{equation}
Accordingly,
$\boldsymbol{\delta}
=\{\pm\mathbf a_{1,2,3},\pm\mathbf b_{1,2,3}\}$,
with $t_{\boldsymbol{\delta}}=t_1$ for
$\boldsymbol{\delta}=\pm\mathbf a_{1,2,3}$ and
$t_{\boldsymbol{\delta}}=t_2$ for
$\boldsymbol{\delta}=\pm\mathbf b_{1,2,3}$.
We set $t_1=1$ as the unit of energy.

In the fully polarized background, the bare hole dispersion is
\begin{equation}
\epsilon_{\mathbf k}
=
-2t_1\sum_{\nu=1}^{3}
\cos(\mathbf k\cdot\mathbf a_\nu)
-
2t_2\sum_{\nu=1}^{3}
\cos(\mathbf k\cdot\mathbf b_\nu).
\label{eq:disp}
\end{equation}
The FM ground state corresponds to $\mathbf k=\Gamma$, with energy
\begin{equation}
E_{\mathrm{FM}}
=
\epsilon_{\Gamma}
=
-6t_1-6t_2.
\end{equation}

For the finite clusters used throughout, we label each lattice
position as
\begin{equation}
\mathbf R_{x,y}=x\mathbf a_1+y\mathbf a_2,
\end{equation}
where $x=0,\ldots,L_x-1$ and $y=0,\ldots,L_y-1$.
For the numerical implementation, these positions are mapped
to a one-dimensional site index using the row-major convention
\begin{equation}
\operatorname{site}(x,y)=x+L_x y+1.
\end{equation}
This one-based indexing is used in both the exact-diagonalization
and sparse-linear-algebra codes. Periodic boundary conditions are imposed
along both primitive lattice directions, so that the coordinates
$x$ and $y$ are understood modulo $L_x$ and $L_y$, respectively.

The projector $\mathcal{P}$ enforces the no-double-occupancy constraint
appropriate to $U=\infty$. Thus, the local physical Hilbert
space at each site is restricted to the three states
$\{|0\rangle,|\uparrow\rangle,|\downarrow\rangle\}$, and
the projection in Eq.~\eqref{eq:H} eliminates every hopping
process that would produce double occupancy. In the one-hole sector,
$N_{\mathrm{el}}=N-1$, so exactly one site is empty and
all remaining sites are singly occupied.

For the Fock basis, we encode each many-body state by two
$N$-bit strings, $b_\uparrow$ and $b_\downarrow$, which
record the occupations of the $\uparrow$ and $\downarrow$ electrons,
respectively. The $(i-1)$th bit of $b_\sigma$ equals 1
if site $i$ is occupied by an electron with spin $\sigma$ and 0 otherwise.
The no-double-occupancy constraint is expressed as the bitwise condition
\begin{equation}
b_\uparrow\,\&\,b_\downarrow=0,
\end{equation}
while the one-hole sector satisfies
\begin{equation}
\operatorname{popcount}(b_\uparrow)
+
\operatorname{popcount}(b_\downarrow)
=
N-1.
\end{equation}
Equivalently, among the lowest $N$ bits, the unique zero
bit of the bitwise union $b_\uparrow\vee b_\downarrow$ identifies the
position of the hole.

\section{Gauge rotation and spiral frame}\label{sec:gauge}

We begin with a conical spiral of wavevector $\mathbf Q$
and polar angle $\theta$. The local spin direction at position
$\mathbf R_j$ is
\begin{equation}
\mathbf n_j
=
\left(
\sin\theta\cos\phi_j,\,
\sin\theta\sin\phi_j,\,
\cos\theta
\right),
\qquad
\phi_j=\mathbf Q\cdot\mathbf R_j.
\end{equation}
The corresponding local majority-spin state is
\begin{equation}
|\uparrow_j\rangle
=
\cos\left(\frac{\theta}{2}\right)|\uparrow\rangle
+
e^{i\phi_j}
\sin\left(\frac{\theta}{2}\right)|\downarrow\rangle.
\end{equation}
Up to an irrelevant site-dependent overall phase, this
spinor is generated by the local rotation
\begin{equation}
\mathcal U_j
=
e^{-i\phi_j\sigma_z/2}
e^{-i\theta\sigma_y/2}.
\end{equation}
This transformation maps the spatially varying spin
configuration onto a uniform ferromagnetic configuration in the local
frame. The hopping between $\mathbf R_j$ and
$\mathbf R_j+\boldsymbol{\delta}$ then acquires the spin-dependent
matrix
\begin{align}
D_{\boldsymbol{\delta}}(\theta)
&
=
\mathcal U_j^\dagger
\mathcal U_{\mathbf R_j+\boldsymbol{\delta}}
\nonumber\\
&
=
e^{i\theta\sigma_y/2}
e^{-i(\mathbf Q\cdot\boldsymbol{\delta})\sigma_z/2}
e^{-i\theta\sigma_y/2}
\nonumber\\
&
=
\begin{pmatrix}
c_{\boldsymbol{\delta}}
-i\cos\theta\,s_{\boldsymbol{\delta}}
&
i\sin\theta\,s_{\boldsymbol{\delta}}
\\[4pt]
i\sin\theta\,s_{\boldsymbol{\delta}}
&
c_{\boldsymbol{\delta}}
+i\cos\theta\,s_{\boldsymbol{\delta}}
\end{pmatrix},
\end{align}
where
\begin{equation}
c_{\boldsymbol{\delta}}
=
\cos\left(\frac{\mathbf Q\cdot\boldsymbol{\delta}}{2}\right),
\qquad
s_{\boldsymbol{\delta}}
=
\sin\left(\frac{\mathbf Q\cdot\boldsymbol{\delta}}{2}\right).
\end{equation}
The coplanar spiral considered in the main text corresponds
to $\theta=\pi/2$. In this limit,
\begin{equation}
D_{\boldsymbol{\delta}}\left(\frac{\pi}{2}\right)
=
\begin{pmatrix}
c_{\boldsymbol{\delta}} & i s_{\boldsymbol{\delta}}
\\[4pt]
i s_{\boldsymbol{\delta}} & c_{\boldsymbol{\delta}}
\end{pmatrix}.
\end{equation}
Thus, the spin-conserving and spin-flip hopping amplitudes
are proportional to
$\cos(\mathbf Q\cdot\boldsymbol{\delta}/2)$ and
$i\sin(\mathbf Q\cdot\boldsymbol{\delta}/2)$, respectively, in agreement
with the convention used below. Reversing the bond orientation,
$\boldsymbol{\delta}\rightarrow-\boldsymbol{\delta}$, changes the sign
of the spin-flip matrix element but leaves the spin-conserving matrix
element unchanged.

The choice $\theta=\pi/2$ is optimal
for the spiral instability
because the spin-flip vertex is proportional to $\sin\theta$, so the
$u\leftrightarrow d$ mixing is maximal for a coplanar
spiral. At the same time, the unperturbed
zero-magnon energy at
$\mathbf k=\Gamma$ is independent of $\theta$.
Indeed, the two spin-conserving contributions are
proportional to
$\epsilon_{\mathbf Q/2}$ and $\epsilon_{-\mathbf Q/2}$ and carry weights
$\cos^2(\theta/2)$ and $\sin^2(\theta/2)$, respectively. Inversion
symmetry implies
$\epsilon_{\mathbf Q/2}=\epsilon_{-\mathbf Q/2}$, while the two weights
sum to unity.
Thus, tilting away from $\theta=\pi/2$ suppresses the
hole--magnon hybridization without lowering the
zero-magnon reference energy.

The \textbf{$uu$ channel} describes the hole hopping to
an empty site while the magnon (flipped spin) remains at its
relative displacement $\mathbf r$ from the hole:
$|\mathbf r\rangle\to|\mathbf r-\boldsymbol{\delta}\rangle$
(for $\mathbf r\neq\boldsymbol{\delta}$).
The \textbf{$dd$ channel} acts when the hole hops onto
the magnon site
($\mathbf r=\boldsymbol{\delta}$): the magnon is displaced to the
hole's former position, giving
$|\boldsymbol{\delta}\rangle\to|-\boldsymbol{\delta}\rangle$.
The \textbf{$ud$ channel} creates or annihilates a
magnon, connecting the zero-magnon state
$|0\text{-mag}\rangle$ to one-magnon relative-coordinate
states $|\mathbf r\rangle$.

\section{Ferromagnetic ground state in the classical limit}\label{sec:convex}

We prove that, within the family of classical
single-$\mathbf Q$ conical spirals, the FM always gives the lowest
single-hole energy for \emph{arbitrary} $t_1$ and $t_2$.

\textbf{Theorem.} For the $J=0$ $t$-$J$ model with a single hole in
a classical single-$\mathbf Q$ conical-spiral
background with wavevector $\mathbf Q$ and
polar angle $\theta$,
\begin{equation}
E_{\mathrm{spiral}}
=
\min_{\mathbf k}
\epsilon(\mathbf k;\mathbf Q,\theta)
\geq
E_{\mathrm{FM}}
=
\min_{\mathbf k}\epsilon_{\mathbf k}
\label{eq:classical_bound}
\end{equation}
for all $\mathbf Q$ and $\theta$.

\textbf{Proof.} In the gauge-rotated frame, the
zero-magnon dispersion in a spiral background is a
convex combination of the bare FM dispersion at two
shifted momenta:
\begin{equation}
\epsilon(\mathbf k;\mathbf Q,\theta)
=
\cos^2\!\left(\frac{\theta}{2}\right)
\epsilon_{\mathbf k-\mathbf Q/2}
+
\sin^2\!\left(\frac{\theta}{2}\right)
\epsilon_{\mathbf k+\mathbf Q/2}.
\label{eq:convex}
\end{equation}
Since
$\cos^2(\theta/2)+\sin^2(\theta/2)=1$, with both coefficients
non-negative, and
$\epsilon_{\mathbf p}\geq E_{\mathrm{FM}}$
for every momentum $\mathbf p$, it follows that
$\epsilon(\mathbf k;\mathbf Q,\theta)\geq E_{\mathrm{FM}}$
for all $\mathbf k$. Taking the minimum over $\mathbf k$ therefore gives
$E_{\mathrm{spiral}}\geq E_{\mathrm{FM}}$.
$\square$

The physical content of this result is that
the unit spin overlap of the FM minimizes the lowest
single-hole energy. Any single-$\mathbf Q$ spiral
replaces the bare dispersion by a convex combination
of momentum-shifted copies of the same dispersion,
which can only raise or leave unchanged the band
minimum. Equivalently, gauge-transforming a spiral to the FM frame
produces an average of the bare dispersion evaluated at
shifted momenta, and such an average cannot lie below
the global band minimum.

This theorem holds for \emph{arbitrary} $t_1$ and $t_2$
and, more generally, for arbitrary translationally
invariant hopping amplitudes on any lattice; it does not require
positive hopping or any sign condition. It establishes that the spiral
instability is necessarily a \emph{quantum} effect:
a classical single-$\mathbf Q$ spiral cannot have a
lower single-hole energy than the FM. Only the quantum dressing
arising from hybridization with magnon excitations,
encoded in the positive coefficient $\Pi_0$, can
overcome the classical energy deficit and drive the transition.

\section{Multi-magnon power counting}\label{sec:multimagnon}

\subsection{Power-counting argument}

Sectors with two or more magnons do not contribute to the stiffness at
order $Q^2$. An $n$-magnon
creation process starting from the zero-magnon sector
requires $n$ spin-flip vertices, and each such vertex contributes one
factor of
$\sin(\mathbf Q\cdot\boldsymbol{\delta}/2)$. Therefore, the
leading $n$-magnon creation amplitude scales as
\begin{equation}
\Gamma^{(n)}
\propto
\prod_{\ell=1}^{n}
\sin\left(
\frac{\mathbf Q\cdot\boldsymbol{\delta}_{\ell}}{2}
\right)
\sim Q^n.
\end{equation}
The associated second-order energy correction is
\begin{equation}
\delta E^{(n)}
\sim
\frac{|\Gamma^{(n)}|^2}{\Delta_n}
=
O(Q^{2n}),
\end{equation}
where $\Delta_n$ is the excitation denominator in the $n$-magnon
sector, which remains of order $Q^0$ in the small-$Q$
expansion relevant here. For $n\geq2$, the contribution starts at
$O(Q^4)$ and therefore cannot modify the coefficient of $Q^2$ in
Eq.~\eqref{eq:Landau_SM}.

\subsection{Numerical confirmation via Trugman-0123}
To validate the power counting
numerically and nonperturbatively within the chosen
variational spaces, we evaluate the spiral stiffness in extended
variational Hilbert spaces that explicitly include
two- and three-magnon states. We construct a nested
family of path-expansion truncations on a $100\times100$ lattice,
using Trugman-style variational
ans\"atze~\cite{Trugman1988} generated by repeated application of the
hopping Hamiltonian to the bare-hole state. These spaces are indexed
by the magnon-number cutoff $n_{\max}$ and the spatial dressing radius
$r$ around the hole:
\begin{itemize}
\item \textbf{Trugman-01}: complete
zero- and one-magnon Hilbert space on the
$100\times100$ lattice with no spatial truncation.
\item \textbf{Trugman-012}:
zero-, one-, and two-magnon sectors, with
the two-magnon dressing restricted to radius $r$
around the hole.
\item \textbf{Trugman-0123}:
zero- through three-magnon sectors, with radius
$r=2$, which is the most extensive multi-magnon
truncation used in this work.
\end{itemize}
For each truncation, we extract the spiral stiffness from the small-$Q$
behavior of $[E(Q)-E_{\mathrm{FM}}]/Q^2$ along $\Gamma$--$M$. The
extracted critical points are
\begin{align*}
\text{Trugman-01 ($Q\to0$, $100^2$):}\quad
& t_{2,c}/t_1=-0.182,\\
\text{Trugman-0123 ($r=2$, $100^2$):}\quad
& t_{2,c}/t_1=-0.186.
\end{align*}
These results are close to the
analytic value $t_{2,c}/t_1=-0.182$ derived from the
zero- and one-magnon Lippmann--Schwinger reduction in
Sec.~\ref{SM:Tmatrix}. The residual shift of approximately $0.004$ in
the Trugman-0123 result is consistent with the finite dressing radius
$r=2$ rather than with a genuine
two- or three-magnon correction to the $Q^2$
coefficient: increasing $r$ further drives this shift toward zero.
The two- and three-magnon sectors therefore
first modify the spiral energy at $O(Q^4)$ and do not
renormalize the spiral stiffness. This
numerically confirms the perturbative power counting
and demonstrates that the analytic result
$\Pi_0=0.170$ captures the full leading-order quantum softening.

\section{Hard-core constraint}
\label{SM:Tmatrix}

In this section, we derive the exact expression for the quantum
reduction $\Pi_0$ of the spiral
stiffness, treating the hard-core hole--magnon
constraint exactly. A compact form is obtained
from the Lippmann--Schwinger equation by projection
onto the two-dimensional bond subspace
$\{|u_1\rangle,|u_2\rangle\}$.

As discussed in the main text, the hybridization vertex
$M_{\mathbf q}
(\mathbf k,\mathbf Q)$
is linear in $Q$:
\begin{equation}
M_{\mathbf q}(\mathbf k,\mathbf Q)=O(Q).
\end{equation}
Therefore, the correction to the ground-state energy
at order $Q^2$ originates entirely from virtual transitions between
the zero-magnon and one-magnon sectors.
States containing two or more magnons first contribute
at $O(Q^4)$ and can therefore be neglected when determining
the spiral stiffness and the critical point at which
the FM first becomes unstable.

Consequently, the calculation of $\Pi_0$ reduces to an
effective two-body problem of a single hole interacting with a single
magnon, with their mutual exclusion enforced by the
hard-core constraint.

\subsection{Relative-coordinate representation}

We work in the sector of zero total momentum,
$\bm K=0$, which contains the ferromagnetic ground state.
The one-hole plus one-magnon states are
\begin{equation}
|{\bm r}\rangle
=
\frac{1}{\sqrt N}
\sum_\ell
h^\dagger_{{\bm R}_\ell+{\bm r}}
a^\dagger_{{\bm R}_\ell}
|{\rm FM}\rangle,
\qquad
{\bm r}\neq 0,
\end{equation}
where $\bm r$ is the displacement of the hole
relative to the magnon.
To treat the hybridization with the zero-magnon state
on equal footing, we associate the excluded
relative coordinate $\bm r=0$ with the physical zero-magnon state
\begin{equation}
|{\bm 0}\rangle
\equiv
\frac{1}{\sqrt N}
\sum_\ell
h^\dagger_{{\bm R}_\ell}|{\rm FM}\rangle .
\end{equation}
This state has zero total momentum and energy
$E_{\rm FM}=\epsilon_{\bm k=0}=-6t_1-6t_2$, as given in
Sec.~\ref{sm:model}. The construction does not introduce an additional
physical state; it is a one-to-one relabeling of the complete
zero- and one-magnon Hilbert space.

We consider an infinitesimal spiral and retain only
the terms required to determine the coefficient of $Q^2$.
The zero-magnon energy has the expansion
\begin{equation}
\epsilon_0(\bm Q)
=
E_{\rm FM}
+
\alpha_{\rm cl}Q^2
+
\mathcal{O}(Q^4),
\end{equation}
where $\alpha_{\rm cl}Q^2$ is the classical contribution to the
spiral stiffness and will be treated separately. The hybridization
between the zero- and one-magnon sectors is instead
$\mathcal{O}(Q)$ and therefore produces an
$\mathcal{O}(Q^2)$ energy correction in second-order perturbation
theory. By contrast, the $Q$ dependence of the one-magnon block
starts at $\mathcal{O}(Q^2)$ and, when inserted between two
$\mathcal{O}(Q)$ hybridization vertices, contributes only at
$\mathcal{O}(Q^4)$. Consequently, in the calculation of $\Pi_0$
both diagonal sectors can be evaluated at $Q=0$, and the only
$Q$ dependence that must be retained is that of the
zero--one-magnon hybridization.
\begin{equation}
\hat H
=
\hat H_0+\hat V,
\end{equation}
with
\begin{equation}
\hat H_0
=
\sum_{\bm k}
\epsilon_{\bm k}
|\bm k\rangle
\langle \bm k|,
\end{equation}
where $\epsilon_{\bm k}$ is the hole dispersion
in the ferromagnetic background, and
\begin{equation}
\hat V
=
E_{\rm FM}
|{\bm 0}\rangle\langle{\bm 0}|
+
\sum_{\bm\delta}
\left[
\tau_{\bm\delta}
|\bm\delta\rangle\langle {\bm 0}|
+
\tau_{\bm\delta}^{*}
|{\bm 0}\rangle\langle\bm\delta|
-
t_{\bm\delta}
|\bm\delta\rangle\langle-\bm\delta|
\right].
\end{equation}
The coefficients
\begin{equation}
\tau_{\bm\delta}
=
t_{\bm\delta}
\left[
1
-
i
\sin\left(
\frac{\bm Q\cdot\bm\delta}{2}
\right)
\right]
\end{equation}
combine the cancellation of the artificial hopping
through $\bm r=0$ generated by $\hat H_0$ with the physical
zero--one-magnon hybridization. The final term in $\hat V$
describes the $dd$ process that exchanges the positions of the
hole and magnon across the bond $\bm\delta$.

\subsection{Projection onto the interaction subspace}

The two-body eigenvalue problem can be written in resolvent form as
\begin{equation}
|\Psi\rangle
=
G_0(E)\,\hat V\,|\Psi\rangle,
\qquad
G_0(E)=\frac{1}{E-\hat H_0},
\label{eq:resolvent}
\end{equation}
which follows directly from the Schrödinger equation
$(\hat H_0+\hat V)|\Psi\rangle=E|\Psi\rangle$.

Although the exact eigenstate $|\Psi\rangle$ has support on
all relative coordinates ${\bm r}$, the interaction
$\hat V$ is of finite range and acts only within
the subspace
\begin{equation}
{\cal S}
=
\{
|{\bm0}\rangle,
|\bm\delta\rangle
\},
\end{equation}
consisting of the auxiliary state $|\bm0\rangle$,
which represents the physical zero-magnon state, and the six nearest-
and six next-nearest-neighbor bond states. Equivalently,
if $P_{\cal S}$ denotes the projector onto this 13-dimensional
subspace, then
\begin{equation}
\hat V=P_{\cal S}\hat V P_{\cal S}.
\end{equation}
Consequently, the full eigenstate is completely
determined by the finite set of amplitudes
\begin{equation}
\psi_\alpha
=
\langle\alpha|\Psi\rangle,
\qquad
\alpha\in{\cal S}.
\end{equation}
Once these amplitudes are known, the wave function
at every relative coordinate outside ${\cal S}$ is reconstructed
from Eq.~(\ref{eq:resolvent}).

Projecting Eq.~(\ref{eq:resolvent}) onto the basis states
$\alpha\in{\cal S}$ yields
\begin{equation}
\psi_\alpha
=
\sum_{\beta,\gamma\in{\cal S}}
\langle\alpha|G_0(E)|\beta\rangle
V_{\beta\gamma}
\psi_\gamma ,
\label{eq:finite_system}
\end{equation}
where
\(
V_{\beta\gamma}
=
\langle\beta|\hat V|\gamma\rangle
\).
Equation~(\ref{eq:finite_system}) therefore reduces
the exact infinite-lattice eigenvalue problem to a homogeneous
linear system of dimension 13.

\subsection{Reduction to the hybridizing sector}

In the limit $Q\rightarrow0$, the hybridization couples
$|{\bm0}\rangle$ only to odd-parity bond combinations.
We therefore introduce
\begin{eqnarray}
|u_1\rangle
&=&
\sum_{\mu=1}^{3}
(\hat{\bm Q}\!\cdot\!\bm a_\mu)
\left(
|\bm a_\mu\rangle
-|-\bm a_\mu\rangle
\right),
\\
|u_2\rangle
&=&
\sum_{\mu=1}^{3}
(\hat{\bm Q}\!\cdot\!\bm b_\mu)
\left(
|\bm b_\mu\rangle
-|-\bm b_\mu\rangle
\right).
\end{eqnarray}
For convenience, we define the corresponding
normalized bond states
\begin{equation}
|\bar u_m\rangle
=
\frac{|u_m\rangle}
{\sqrt{\langle u_m|u_m\rangle}},
\qquad
m=1,2,
\end{equation}
together with the hopping vector
\begin{equation}
{\bm t}
=
\begin{pmatrix}
t_1\sqrt{\langle u_1|u_1\rangle}\\
t_2\sqrt{\langle u_2|u_2\rangle}
\end{pmatrix}.
\end{equation}
To leading order in $Q$, the one-magnon state
generated by the hybridization is
\begin{equation}
|\Gamma(Q)\rangle
=
-\frac{iQ}{2}
\sum_{m=1}^{2}
({\bm t})_m
|\bar u_m\rangle
+
\mathcal{O}(Q^3).
\end{equation}

Thus, only the two-dimensional bond subspace
spanned by $|\bar u_1\rangle$ and $|\bar u_2\rangle$ couples
directly to the zero-magnon state at order $Q$. The full
one-magnon wave function, however, is not confined to these bond
states, because propagation generates amplitudes at arbitrary
relative coordinates. We therefore decompose the eigenstate as
\begin{equation}
|\Psi\rangle
=
A_0|{\bm0}\rangle
+
|\psi_1\rangle,
\end{equation}
where $|\psi_1\rangle$ denotes the complete
one-magnon component. Let $\hat H_{\rm 1m}$ be the one-magnon
Hamiltonian evaluated at $Q=0$, including the hard-core
hole--magnon scattering, and define its exact resolvent by
\begin{equation}
G(E)
=
\frac{1}{E-\hat H_{\rm 1m}}.
\end{equation}

Projecting the Schrödinger equation onto the
zero-magnon sector gives
\begin{equation}
\left[E-\epsilon_0(Q)\right]A_0
=
\langle\Gamma(Q)|\psi_1\rangle,
\end{equation}
whereas projection onto the one-magnon sector gives
\begin{equation}
\left(E-\hat H_{\rm 1m}\right)|\psi_1\rangle
=
|\Gamma(Q)\rangle A_0.
\end{equation}
The latter equation implies
\begin{equation}
|\psi_1\rangle
=
G(E)|\Gamma(Q)\rangle A_0.
\end{equation}
Eliminating $|\psi_1\rangle$ yields the exact
secular equation
\begin{equation}
E-\epsilon_0(Q)-\Sigma(E,Q)=0,
\end{equation}
with
\begin{equation}
\Sigma(E,Q)
=
\langle\Gamma(Q)|
G(E)
|\Gamma(Q)\rangle.
\end{equation}

Because $|\Gamma(Q)\rangle$ lies entirely in the
two-dimensional hybridizing sector, only the projection of the
exact resolvent onto this subspace is required. We define
\begin{equation}
{\cal G}_{mn}(E)
=
\langle\bar u_m|
G(E)
|\bar u_n\rangle,
\qquad
m,n=1,2.
\end{equation}
Within this invariant odd-parity sector, the Dyson
equation reduces to
\begin{equation}
{\cal G}(E)
=
\left[
{\cal G}^{0}(E)^{-1}
-
W
\right]^{-1},
\end{equation}
where ${\cal G}^{0}(E)$ is the projected vacancy
resolvent obtained without the $dd$ interaction, and $W$ is the
$dd$ interaction projected onto the normalized bond basis.

Using the leading-order form of
$|\Gamma(Q)\rangle$, the self-energy becomes
\begin{equation}
\Sigma(E,Q)
=
\frac14
Q^2
{\bm t}^{T}
{\cal G}(E)
{\bm t}
+
\mathcal{O}(Q^4).
\end{equation}
Since the prefactor is already of order $Q^2$,
the resolvent can be evaluated at $E=E_{\rm FM}$. Using
\begin{equation}
\epsilon_0(Q)
=
E_{\rm FM}
+
\alpha_{\rm cl}Q^2
+
\mathcal{O}(Q^4),
\end{equation}
the secular equation gives
\begin{equation}
E(Q)-E_{\rm FM}
=
\left[
\alpha_{\rm cl}
+
\frac14
{\bm t}^{T}
{\cal G}(E_{\rm FM})
{\bm t}
\right]Q^2
+
\mathcal{O}(Q^4).
\end{equation}
Comparing with Eq.~(\ref{eq:EQ}) of the main text yields
\begin{equation}
\Pi_0
=
-\frac14
{\bm t}^{T}
{\cal G}(E_{\rm FM})
{\bm t},
\end{equation}
which is the exact quantum correction to the spiral
stiffness.

\section{Order of the FM-to-spiral transition}\label{sec:transition_order}

The sign of the quartic coefficient $c_4$ at the
point where the quadratic stiffness vanishes determines whether
the ferromagnetic instability develops continuously or is preempted
by a first-order transition. The spiral energy has the Landau
expansion
\begin{equation}
E(Q) - E_\mathrm{FM}
=
(\alpha_\mathrm{cl} - \Pi_0)\,Q^2
+
c_4\,Q^4
+
\mathcal{O}(Q^6),
\label{eq:Landau_SM}
\end{equation}
with $\alpha_\mathrm{cl} = \Pi_0$ at
$t_2=t_{2,c}$. If $c_4>0$, minimization on the
spiral side gives
\begin{equation}
Q^*
=
\sqrt{\frac{\Pi_0-\alpha_{\rm cl}}{2c_4}}
\propto
\sqrt{|t_2-t_{2,c}|},
\end{equation}
so the optimal wavevector grows continuously from
zero and the transition at $t_{2,c}$ is continuous. If $c_4<0$,
stability must instead be restored by higher-order terms in $Q$.
A minimum at finite $Q$ then appears before the quadratic stiffness
vanishes, preempting the Lifshitz instability and producing a
first-order transition.

\subsection{Three-term decomposition by magnon number}

Because each spin-flip vertex changes the magnon
number by one and scales as $M_{\bm q}\sim Q$, reaching an
$n$-magnon intermediate sector from the zero-magnon state and
returning to it requires at least $2n$ such vertices. The leading
contribution from that sector therefore scales as $Q^{2n}$.
The $Q^4$ coefficient is consequently determined
entirely by the zero-, one-, and two-magnon sectors; sectors
containing three or more magnons first contribute at
$\mathcal{O}(Q^6)$. It can thus be decomposed into three physically
distinct contributions:
\begin{equation}
c_4
=
-\beta_\mathrm{cl}
+
c_4^{\mathrm{1\text{-}mag}}
+
c_4^{\mathrm{2\text{-}mag}}.
\label{eq:c4_SM}
\end{equation}

\textbf{Classical quartic stiffness.}
Expanding $\cos(\bm Q{\cdot}\bm\delta/2)$ to fourth order in $Q$
and summing over the NN and NNN bonds for
$\bm Q$ along the $\Gamma$--$M$ direction yields
\begin{equation}
-\beta_\mathrm{cl}
=
-\frac{3t_1+27t_2}{512},
\label{eq:beta_cl}
\end{equation}
equivalent to the unsimplified form
$-(9t_1+81t_2)/1536$.
At $t_2=t_{2,c}=-0.182\,t_1$, this gives
$-\beta_\mathrm{cl}\approx+0.0037\,t_1$, a small positive
contribution.

\textbf{One-magnon (polaronic) correction.}
The contribution
$c_4^{\mathrm{1\text{-}mag}}$ contains all terms of order $Q^4$
generated within the zero- and one-magnon sectors. It is the
quartic counterpart of the same hole self-energy that produces
$\Pi_0$ at order $Q^2$ and includes the subleading $Q$ dependence
of both the hybridization vertex and the one-magnon resolvent.
The Trugman-01 $Q$ scan on a $100\times100$ lattice yields
\begin{equation}
c_4^{\mathrm{1\text{-}mag}}
\approx
+0.015\,t_1.
\end{equation}
The combined classical and one-magnon contribution is therefore
\begin{equation}
-\beta_\mathrm{cl}
+
c_4^{\mathrm{1\text{-}mag}}
\approx
+0.019\,t_1,
\end{equation}
which is the quartic coefficient obtained from the Trugman-01
calculation alone.

\textbf{Two-magnon correction.}
The first contribution involving the two-magnon
sector arises from the sequence
$0\rightarrow1\rightarrow2\rightarrow1\rightarrow0$, which
contains four spin-flip vertices and is therefore of order $Q^4$.
In spin-wave language, this process corresponds to a virtual
magnon-pair fluctuation generated by an effective anomalous
amplitude $B_{\bm q}\sim Q^2$. The associated zero-point
correction consequently has the form
\begin{equation}
E_{\rm zp}(\bm Q)
=
-\chi Q^4
+
\mathcal{O}(Q^6),
\qquad
\chi>0.
\label{eq:Ezp_SM}
\end{equation}
This term lowers the energy, so that
$c_4^{\mathrm{2\text{-}mag}}=-\chi<0$, and is absent from any
one-magnon calculation by construction. Because it begins at
$\mathcal{O}(Q^4)$, it does not renormalize the quadratic
coefficient $\Pi_0$ or shift the point at which the ferromagnetic
stiffness vanishes. This is consistent with the exact
Lippmann--Schwinger calculation of Sec.~\ref{SM:Tmatrix}, which
involves only the zero- and one-magnon sectors.

\subsection{Two-magnon contribution from Trugman-012}
\label{sec:c4_trugman012}

The two-magnon contribution
$c_4^{\mathrm{2\text{-}mag}}$ is evaluated numerically using the
Trugman-012 diagonalization. This calculation
constructs the complete zero-, one-, and two-magnon variational
space at fixed total momentum and diagonalizes the resulting
Hamiltonian as a single sparse-matrix problem. Calculations were
performed on lattices as large as $30\times30$, with the extracted
quartic coefficient already stable between $12\times12$ and
$14\times14$.
The same construction was used in Fig.~\ref{fig:SM_polaron} on the
$3\sqrt{3}\times3\sqrt{3}$ cluster, where the Hilbert-space
dimension is $352$, to test the variational polaron hierarchy.

To isolate the two-magnon contribution, we extract
$c_4$ from Trugman-01 and Trugman-012 calculations performed on
the same lattice and using the same small-$Q$ fitting window.
Because Trugman-012 reproduces the quadratic coefficient $\Pi_0$
and contains the complete lower-magnon sectors, the incremental
quartic shift produced by allowing two magnons is
\begin{equation}
c_4^{\mathrm{2\text{-}mag}}
=
c_4^{(012)}
-
c_4^{(0+1)}
\approx
-0.011\,t_1.
\label{eq:c4_2mag_result}
\end{equation}
This is a sizable negative two-magnon contribution. It reduces,
but does not overturn, the positive classical and one-magnon
terms. Combining the three pieces gives
\begin{equation}
c_4
=
-\beta_\mathrm{cl}
+
c_4^{\mathrm{1\text{-}mag}}
+
c_4^{\mathrm{2\text{-}mag}}
\approx
+0.006\ \text{to}\ +0.008\,t_1
>0.
\end{equation}
The range reflects the difference between
$c_4^{(0+1)}\approx+0.017\,t_1$ obtained from the same-lattice
$12\times12$ extraction and the converged value
$c_4^{(0+1)}\approx+0.019\,t_1$ obtained from the
$100\times100$ Trugman-01 scan. In either case, the total quartic
coefficient remains positive, establishing that the
ferromagnetic-to-spiral transition is continuous.

\textbf{Comparison with the linear-spin-wave estimate.}
A Holstein--Primakoff linear-spin-wave treatment of the
rotated-frame Hamiltonian, retaining only bilinear magnon terms
and replacing $\sqrt{2S-n}$ by $\sqrt{2S}$, neglects both
hard-core constraints. It gives the bosonic Bogoliubov
zero-point correction
\begin{equation}
E_{\rm zp}(\bm Q)
=
\frac{1}{2}
\sum_{\bm q}
\left[
\omega_{\bm q}(\bm Q)
-
\omega_{\bm q}^{(\rm FM)}
\right]
=
-\chi Q^4
+
\mathcal{O}(Q^6),
\end{equation}
with $\chi/t_1\simeq2.5\times10^{-2}$ at
$t_2=t_{2,c}$. The resulting estimate,
\begin{equation}
c_4
\approx
+0.0037+0.015-0.025
\approx
-0.006,
\end{equation}
would incorrectly place the system on the first-order side.

The exact Trugman-012 result,
$c_4^{\mathrm{2\text{-}mag}}\approx-0.011\,t_1$, is of the same
order as the linear-spin-wave estimate but approximately a factor
of two smaller in magnitude. This reduction
reflects the two local constraints absent from linear spin-wave
theory: the hole and a magnon cannot occupy the same site, and
two magnons cannot occupy the same site because the local
spin-$1/2$ Hilbert space permits at most one spin flip.
Enforcing these constraints reduces
$|c_4^{\mathrm{2\text{-}mag}}|$ from approximately $0.025\,t_1$
to $0.011\,t_1$, changing the total quartic coefficient from
approximately $-0.006\,t_1$ to $+0.006$--$+0.008\,t_1$.
The hard-core constraints therefore suppress,
but do not eliminate, the negative two-magnon correction. This
suppression is sufficient for the positive classical and
one-magnon contributions to prevail, making the transition
continuous.

\subsection{Empirical sign of \texorpdfstring{$c_4$}{c4} from ED and variational data}

The sign of $c_4$ established numerically is independently corroborated
by ED and flux-insertion. On every cluster studied (Sec.~\ref{sec:ED})
and across the full flux range (Sec.~\ref{sec:spectral}), the
variationally optimal spiral wavevector is the \emph{smallest}
available BZ mesh point, with no competing finite-$Q$ minimum and no
discontinuous jump in $\bm Q^*$ as $t_2$ crosses $t_{2,c}$. The
FM-to-spiral transition is therefore continuous, with $Q^* \propto
\sqrt{|t_2 - t_{2,c}|}$ near the critical point.

\section{ED phase diagram}\label{sec:ED}

\begin{table}[h]
\caption{FM$\to$non-FM transition $t_{2,c}$ from exact diagonalization
sweeps across all $S_z$ sectors on finite clusters with full bulk NNN
coordination (6 NNN neighbors per site).}\label{tab:ED}
\begin{ruledtabular}
\begin{tabular}{cccc}
Lattice & $N$ & $t_{2,c}$ & Transition type \\
\hline
$4 \times 4$ & 16 & $\approx -0.24$ & Sharp, re-entrant \\
$4 \times 5$ & 20 & $\approx -0.22$ & Cascade \\
$4 \times 6$ & 24 & $\approx -0.20$ & Intermediate $S$ \\
$5 \times 5$ & 25 & $\approx -0.22$ & Intermediate $S$ \\
$3\sqrt{3}{\times}3\sqrt{3}$ & 27 & $\approx -0.21$
  & Gradual cascade \\
$4 \times 7$ & 28 & $\approx -0.20$ & Sharp \\
$5 \times 6$ & 30 & $\approx -0.21$ & Intermediate $S$ \\
\end{tabular}
\end{ruledtabular}
\end{table}

Table~\ref{tab:ED} summarizes the ED results on seven cluster
geometries with the correct bulk NNN
coordination~\cite{Hanisch1997}, computed via momentum-sector
defect-basis ED with GPU acceleration.  The finite-size $t_{2,c}$
values converge to $\approx -0.20$ to $-0.21$, consistent with the
dressed one-loop $t_{2,c} = -0.182$ plus finite-size corrections of
order $\sim 1/(L^2 \ln L)$ from the 2D logarithmic vacancy Green's
function.

The transition type varies with geometry: some lattices show a sharp
FM$\to$low-$S$ jump (4$\times$7), while others exhibit a gradual spin
cascade through intermediate $S$ values
($3\sqrt{3}{\times}3\sqrt{3}$: $S = 13 \to 11 \to 7 \to 1 \to 0$ at
successive crossings near $t_2 \approx -0.170$, $-0.188$, $-0.205$,
and $-0.209$).
These differences reflect finite-size commensurability
effects~\cite{Haerter2005,Sposetti2014} and do
not persist in the thermodynamic limit where the transition is
continuous ($Q \to 0$).

\begin{figure}[t]
\includegraphics[width=\columnwidth]{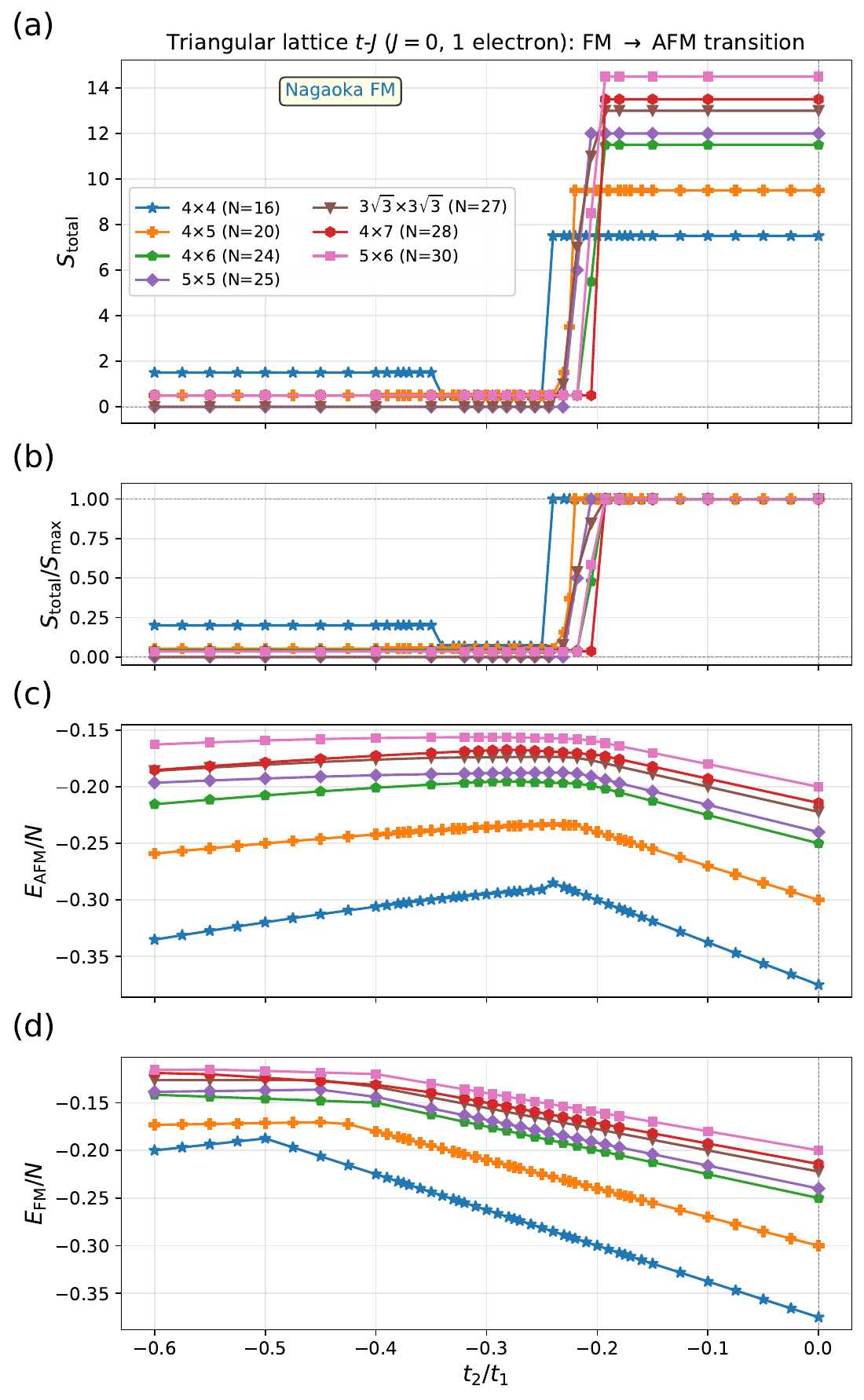}
\caption{ED phase diagram.  (a)~Total spin $S$ and (b)~normalized
$S/S_\mathrm{max}$ vs~$t_2$ on seven clusters with $L_x \geq 4$.
The FM ($S = S_\mathrm{max}$) is destabilized at
$t_{2,c}^\mathrm{ED} \approx -0.20$ to $-0.22$.
(c,d)~Ground state energy per site from the AFM ($S=0$ or $1/2$)
and FM sectors, showing the level crossing at the
transition.}\label{fig:SM_phase}
\end{figure}

\begin{figure}[t]
\includegraphics[width=\columnwidth]{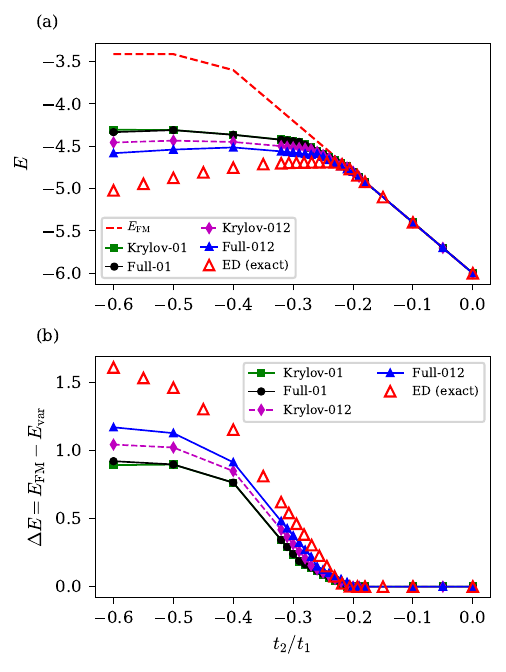}
\caption{Variational spin polaron on $3\sqrt{3}{\times}3\sqrt{3}$
($N = 27$).  (a)~Ground state energy from four variational levels,
Krylov-01 (0+1 magnon, dim\,=\,13), Full-01 (dim\,=\,27), Krylov-012
(0+1+2 magnon, dim\,=\,79), Full-012 (dim\,=\,352), compared with
exact ED (red triangles). 
(b)~Energy gain $\Delta E = E_\mathrm{FM}
- E_\mathrm{var}$ quantifying the quantum advantage of each
variational level over the Nagaoka FM\@.  The Full-012 captures
70--85\% of the exact $\Delta E$ in the spiral
regime~\cite{SSZhang2018,Davydova2023}. }
\label{fig:SM_polaron}
\end{figure}

\section{Spectral flow and ground state identification}\label{sec:spectral}

\subsection{Spectral-flow method}

For the larger clusters, we track the low-energy levels by spectral
flow under flux insertion. A charge flux $\Phi$ is
threaded through one cycle of the torus by imposing the twisted
boundary condition
\begin{equation}
c_{{\bm r}+{\bm L},\sigma}
=
e^{i\Phi}
c_{{\bm r},\sigma}.
\end{equation}
Equivalently, the hopping amplitudes crossing the
corresponding boundary acquire a Peierls phase $e^{i\Phi}$. In a
uniform gauge, the twist continuously shifts the allowed crystal
momenta by $\Phi/L$, thereby probing momenta between the discrete
points available at zero flux.

Because the Peierls phase is identical for both spin species,
the flux preserves spin-rotation symmetry and, in
particular, $S_z$ remains a good quantum number throughout the
sweep. In the $N=27$, $S_z=0$ sector, for which
$N_\uparrow=N_\downarrow=13$, of the
$3\sqrt{3}\times3\sqrt{3}$ cluster, each momentum block has
\begin{equation}
\dim\mathcal{H}_{\bm k}
=
\binom{26}{13}
=
10\,400\,600
\simeq
10.4\times10^6.
\end{equation}
These blocks were diagonalized using GPU-accelerated sparse
eigensolver routines.

The continuous momentum interpolation provided by
the flux resolves the evolution of the lowest-energy branch even
when the zero-flux momentum grid is poorly commensurate with the
incipient long-wavelength spiral. In particular, the appearance of
a state with $E(\Phi)<E_{\rm FM}$ at the smallest accessible
nonzero flux directly signals that the ferromagnetic state is
unstable toward a spiral with a small wavevector.

\subsection{Evidence for a $Q\to0$ spiral instability}

At $t_2=-0.200\,t_1$ on the
$3\sqrt{3}\times3\sqrt{3}$ cluster,
$E(\Phi=0)=E_{\rm FM}$ exactly, because the finite cluster still
selects the ferromagnetic state at zero flux. At the smallest
nonzero flux shown, $\Phi=\pi/20$, the energy is already lower than
the ferromagnetic value,
\begin{equation}
E(\pi/20)-E_{\rm FM}
=
-0.004\,t_1.
\end{equation}
Moreover, the spectral-flow branch evolves smoothly
downward from $\Phi=0$. Because the twist continuously shifts the
allowed momentum away from $\Gamma$, this negative small-flux
response shows that the ferromagnetic state becomes unstable as
soon as sufficiently long-wavelength spiral states are made
accessible. Together with the negative $Q^2$ stiffness obtained
analytically and from the Trugman-01 calculation, this identifies
the instability as the $Q\to0$ spiral instability.

Several additional diagnostics support this identification.

\emph{Gradual spin cascade.}
On the $3\sqrt{3}\times3\sqrt{3}$ cluster, the ground-state total
spin evolves as
\begin{equation}
S=13\rightarrow11\rightarrow7\rightarrow1\rightarrow0
\end{equation}
as $t_2$ is varied approximately from $-0.17\,t_1$ to
$-0.25\,t_1$.
This sequence shows that the finite-size ground
state loses its polarization progressively rather than through a
single direct level crossing from $S=S_{\rm max}$ to the
minimum-spin sector. Although the spin cascade alone does not
uniquely determine the nature of the ordered state, it is
consistent with a spiral whose wavevector and magnon content grow
continuously upon moving away from the ferromagnetic boundary.

\emph{Ground-state momentum near the transition.}
On all clusters studied, with $N=16$--$30$, the ground state
immediately below the ferromagnetic boundary has momentum
$\bm k=\Gamma$, rather than $\bm k=K$.
This observation is particularly informative on
the $3\sqrt{3}\times3\sqrt{3}$ cluster, for which the $K$ point
belongs to the allowed momentum mesh. On this cluster, the ground
state remains in the $\Gamma$ sector down to
$t_2\simeq-0.25\,t_1$ and shifts to the $K$ sector only at stronger
frustration. Thus, a $K$-centered $120^\circ$ state is not the
first instability of the Nagaoka ferromagnet. The momentum
evolution instead supports an initially long-wavelength state,
whose incommensurate wavevector is represented indirectly through
finite-size level crossings and spectral flow.

\emph{Absence of a preempting finite-$Q$ instability.}
The full spectral flow shows that the lowest branch
emerges continuously from $\Phi=0$, with no separate finite-flux
minimum crossing below the ferromagnetic state before the
small-$\Phi$ instability. Likewise, the variational energy
$E(Q)$ develops its minimum continuously from $Q=0$, rather than
through a jump to a disconnected finite-$Q$ minimum. Finally, the
positive quartic coefficient
\begin{equation}
c_4
\simeq
+0.006\text{--}+0.008\,t_1
>0
\end{equation}
ensures that the local Landau expansion is stable
when the quadratic stiffness changes sign. These results rule out
a first-order finite-$Q$ state preempting the long-wavelength
instability and establish the spiral as the first instability of
the Nagaoka ferromagnet.

\subsection{Twofold degeneracy and $C_3$ identification}

On the $3\sqrt{3}\times3\sqrt{3}$ torus, the two lowest states in
the $\Gamma$ momentum sector become degenerate
to machine precision for $t_2/t_1\leq-0.25$:

\begin{center}
\begin{tabular}{ccc}
$t_2/t_1$ & $|E_1-E_2|/t_1$ & Interpretation \\
\hline
$-0.200$ & $3.0\times10^{-3}$  & Split pair near the FM boundary \\
$-0.250$ & $2.2\times10^{-14}$ & $E$ co-representation \\
$-0.300$ & $6.2\times10^{-15}$ & $E$ co-representation \\
$-0.400$ & $1.8\times10^{-15}$ & $E$ co-representation \\
\end{tabular}
\end{center}

This degeneracy should not be interpreted as a
literal folding of the incommensurate wavevectors
$\pm\bm Q^*$ onto the nearest point of the finite-size momentum
mesh. A finite torus admits only its discrete crystal momenta.
Instead, the exact degeneracy follows from the transformation
properties of the two states under the $C_3$ point-group symmetry
and time reversal.

The three complex irreducible representations of $C_3$ have
eigenvalues
\begin{equation}
1,\qquad
\omega=e^{2\pi i/3},
\qquad
\omega^*=e^{-2\pi i/3}.
\end{equation}
Suppose that one member of the pair satisfies
\begin{equation}
C_3|\psi_+\rangle
=
\omega|\psi_+\rangle.
\end{equation}
Because the zero-flux Hamiltonian is invariant
under time reversal $\mathcal T$ and
$[\mathcal T,C_3]=0$, its time-reversed partner obeys
\begin{equation}
C_3\mathcal T|\psi_+\rangle
=
\omega^*\mathcal T|\psi_+\rangle.
\end{equation}
The two states therefore have the same energy but
belong to distinct complex-conjugate $C_3$ sectors and are
necessarily orthogonal. The pair forms the two-dimensional
$E$ co-representation of the symmetry group generated by $C_3$
and time reversal.

For the spiral interpretation, the finite-cluster
eigenstates should be viewed as symmetry-adapted combinations of
the $C_3$-related spiral configurations. The emergence of the
$E$ doublet is therefore consistent with a spiral that selects a
direction in the thermodynamic limit, while the symmetry-restored
finite-size eigenstates retain definite $C_3$ quantum numbers.

Direct evaluation using the ED eigenvectors confirms this
identification. The representation of $C_3$ within the degenerate
subspace is obtained from the $2\times2$ matrix
\begin{equation}
M_{ij}
=
\langle\psi_i|C_3|\psi_j\rangle.
\end{equation}
Numerically, it satisfies
\begin{equation}
\left\|M^\dagger M-I\right\|<10^{-14},
\qquad
\left\|M^3-I\right\|<4\times10^{-14},
\end{equation}
and its eigenvalues agree with $\omega$ and
$\omega^*$ to better than $2\times10^{-14}$. This verifies that
the degeneracy is the symmetry-protected $E$ co-representation,
rather than an accidental near-crossing. The absence of the same
doublet on rectangular clusters without $C_3$ symmetry further
supports its point-group origin rather than a topological
degeneracy.

\subsection{Ruling out a quantum spin liquid}

The twofold degeneracy at $\Gamma$ could superficially resemble
the topological ground-state degeneracy of a chiral spin liquid.
However, the symmetry, geometry, and spectral-flow
properties of the doublet demonstrate that its origin is the
$C_3$ point-group symmetry rather than topological order.

First, as shown above, the two degenerate states
have $C_3$ eigenvalues $\omega$ and $\omega^*$ and form the
two-dimensional $E$ co-representation enforced by $C_3$ and time
reversal $\mathcal T$. Their degeneracy therefore follows directly
from an ordinary symmetry of the finite cluster. No topological
mechanism is required.

Second, the doublet occurs only on clusters that
preserve $C_3$. It is absent on the rectangular
$4\times6$, $5\times5$, and $4\times7$ clusters, where the two
complex-conjugate $C_3$ sectors are not defined. This strong
dependence on point-group symmetry is characteristic of a
symmetry-protected doublet. By contrast, a topological
ground-state manifold is associated with the noncontractible
cycles of the torus and should not require a particular point-group
symmetry, although its finite-size splitting may depend on the
cluster geometry.

Third, when the boundary conditions are twisted,
the $E$ doublet splits on the scale of the many-body bandwidth and
the lowest-energy state migrates among different momentum sectors.
Indeed, the ground state lies at $\Gamma$ for only approximately
$10$--$34\%$ of the sampled flux values. The two states therefore
behave as dispersive symmetry-related levels, rather than as an
isolated quasi-degenerate topological manifold whose splitting
would become exponentially small with increasing system size.

Most importantly, this finite-size doublet appears
within a phase that is continuously connected to the analytically
established $Q\to0$ instability of the Nagaoka ferromagnet. The
negative spiral stiffness, the smooth evolution of the optimal
wavevector from $Q=0$, the positive quartic coefficient, and the
spectral flow all provide direct evidence for conventional
symmetry-breaking spiral order. No additional low-energy manifold
or intervening transition indicative of a chiral spin liquid is
observed.

These results rule out a topological interpretation
of the observed twofold degeneracy. The doublet is instead the
$C_3$- and time-reversal-related finite-size precursor of the
incommensurate spiral state.

\section{Quasiparticle decay in the spiral phase}\label{sec:decay}

\begin{figure}[t]
\includegraphics[width=\columnwidth]{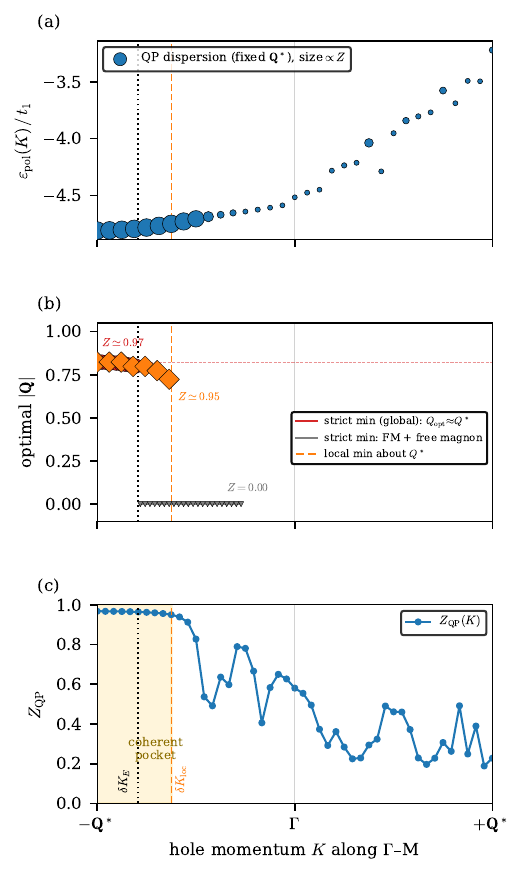}
\caption{Per-momentum spiral stability and quasiparticle decay at
$t_2/t_1=-0.2$ ($24\times24$, one-magnon level). All panels share the
$-\mathbf Q^*\!\to\!\Gamma\!\to\!+\mathbf Q^*$ axis, with
$\delta K=|\mathbf K+\mathbf Q^*|$ and vertical guides at
$\delta K_E=0.170$ and $\delta K_{\rm loc}\simeq0.31$.
(a)~Polaron dispersion on the fixed $\mathbf Q^*$ spiral; marker size
$\propto Z$.
(b)~Optimal $|\mathbf Q|$ at fixed $\mathbf K$; marker size
$\propto Z$. The global minimum remains near
$|\mathbf Q^*|$ until the zero-weight crossing at $\delta K_E$
(red to gray), whereas the locally stable $\mathbf Q^*$ branch
(orange) persists to $\delta K_{\rm loc}$. The finite momentum grid
blocks decay associated with the small
$\mathcal{O}(\delta K^2)$ shift of the optimal wavevector.
(c)~Quasiparticle weight $Z_{\rm QP}(\mathbf K)$:
nearly constant in the protected pocket and
decreasing once local spiral stability is lost at
$\delta K_{\rm loc}$, which defines the decay threshold
$\mathbf K_c$.}\label{fig:SM_decay}
\end{figure}

The polaron dispersion in Fig.~\ref{fig:polaron}(a) of the main text exhibits a
sharp coherent quasiparticle near the band minimum at
$\mathbf K=-\mathbf Q^*$, which dissolves into a multi-magnon
continuum away from it.
We show below that this loss of coherence originates
from the momentum-dependent local stability of the spiral
background. As the quasiparticle center-of-mass momentum is shifted
away from the band minimum, the locally optimal spiral wavevector
changes weakly until the minimum connected to $\mathbf Q^*$ ceases
to exist. Beyond this point, the emission of additional magnons
becomes energetically favorable and the coherent polaron evolves
into the multi-magnon continuum.

To define the threshold unambiguously, we introduce
\begin{equation}
\mathbf K_{\min}=-\mathbf Q^*,
\qquad
\delta\mathbf K
=
\mathbf K-\mathbf K_{\min}
=
\mathbf K+\mathbf Q^*.
\end{equation}
Along the branch shown in the figure, which extends
from $-\mathbf Q^*$ toward increasing momentum, the decay threshold is
\begin{equation}
\mathbf K_c
=
\mathbf K_{\min}
+
\delta\mathbf K_{\rm loc}
=
-\mathbf Q^*
+
\delta\mathbf K_{\rm loc},
\end{equation}
with a symmetry-related threshold on the opposite
side of the quasiparticle minimum.

All results in this section are obtained for $t_2/t_1=-0.2$ on a
$24\times24$ lattice at the one-magnon (C+1m) level. The Hilbert
space consists of the bare hole in a coplanar spiral background and
all one-magnon states and therefore has dimension
\begin{equation}
1+(N-1)=N.
\end{equation}
The Hamiltonian is diagonalized exactly within this space, without
a Lanczos or path-length truncation.

Throughout the calculation, the rotated-frame Bloch momentum
$\mathbf K$ is held fixed while the spiral wavevector $\mathbf Q$
is varied independently. The coplanar tilt is fixed at
$\theta=\pi/2$. At the one-magnon level, the ground-state
wavevector is
\begin{equation}
|\mathbf Q^*|=0.824,
\end{equation}
in agreement with the one-magnon $T$-matrix calculation.
This one-magnon value differs from the fully
optimized multi-magnon value used for the dispersion in the main
text; the purpose of the present calculation is to isolate the
mechanism controlling the loss of local spiral stability.

Fixing $\theta$ is essential. In the limit $\theta\to0$, the spin
configuration becomes a collinear ferromagnet and $\mathbf Q$
enters only as a spin-dependent Peierls phase. It then merely
relabels the Bloch momentum. Allowing the
optimization to move along this gauge direction, or constraining
$\mathbf K$ and $\mathbf Q$ to vary together, would therefore
produce an artificial flattening of the energy and obscure the
physical stability of the coplanar spiral.

\textbf{Optimal spiral at fixed momentum.}
Figure~\ref{fig:SM_decay}(b) shows the optimal spiral wavevector at
each momentum, determined using two complementary criteria.

(i) Global optimization.
We first minimize the exact lowest eigenvalue globally over
$\mathbf Q$. For momentum displacements
$\delta K\leq0.16$ from the band minimum, the optimal wavevector
remains close to its ground-state value, changing only from
$|\mathbf Q|=0.823$ to $0.800$, while the minimizing state remains
highly coherent, with $Z\simeq0.97$.
Reflection symmetry about the band minimum requires
the leading shift of $|\mathbf Q|$ to be even in $\delta K$.
Consequently,
\begin{equation}
|\mathbf Q_{\rm opt}(\delta K)|
-
|\mathbf Q^*|
=
\mathcal{O}(\delta K^2),
\label{eq:DK}
\end{equation}
which explains the weak momentum dependence of the
optimal spiral wavevector near the quasiparticle minimum.

At
\begin{equation}
\delta K_E=0.170,
\end{equation}
the global minimum switches discontinuously to a state with zero
quasiparticle weight. This state consists of a hole
at the ferromagnetic band minimum and a zero-energy magnon carrying
the remaining momentum. At $\mathbf Q=0$, the spin-flip
hybridization vanishes, so this one-magnon state is exactly
decoupled from the bare-hole sector and has $Z=0$. The crossing
therefore changes the global minimum of the variational energy but
does not generate a pole in the single-hole spectral function.
Accordingly, it does not mark the decay threshold of the coherent
polaron.

Since Eq.~\eqref{eq:DK} tells us that $\mathbf Q^*$ is optimal only at
the band minimum,  the quasiparticle can
relax in the thermodynamic limit toward this slightly shifted spiral by emitting a magnon with
momentum of order $|\delta\mathbf Q_{\rm opt}|$. On the finite
$24\times24$ lattice, however, the allowed magnon momenta are
separated by $\Delta k\sim2\pi/L$. As long as
$|\delta\mathbf Q_{\rm opt}|<\Delta k$, no magnon on the finite-size
momentum grid can accommodate this relaxation. The discreteness of
the available momenta therefore kinematically protects the
quasiparticle and accounts for the nearly constant $Z$ observed
close to the band minimum, even though $\mathbf Q^*$ is not exactly
the optimal wavevector at nonzero $\delta K$.

(ii) Local stability of the $\mathbf Q^*$ branch.
To determine whether the spiral supporting the coherent polaron
remains locally stable, we minimize the exact lowest eigenvalue
with respect to $\mathbf Q$ using a local descent initialized at
$\mathbf Q^*$. A local minimum continuously connected to the
ground-state spiral persists, with $Z\simeq0.95$, up to
\begin{equation}
\delta K_{\rm loc}
=
0.31\pm0.01.
\end{equation}
Beyond this displacement, the local barrier separating the
$\mathbf Q^*$ branch from the smaller-$|\mathbf Q|$ scattering
configurations disappears and the descent escapes from the spiral
minimum. Thus, the coherent branch does not soften
continuously toward $\mathbf Q=0$: its optimal wavevector changes
only weakly before the local minimum terminates at
$\delta K_{\rm loc}$.

\textbf{Decay mechanism and threshold.}
Figure~\ref{fig:SM_decay}(c) shows the quasiparticle weight
$Z_{\rm QP}(\mathbf K)$ along the full momentum path and relates
the two momentum scales to the onset of decay. For
$\delta K<\delta K_{\rm loc}$, the $\mathbf Q^*$-connected spiral
is a genuine local minimum and the quasiparticle weight remains
nearly constant,
\begin{equation}
Z_{\rm QP}\simeq0.95\text{--}0.97.
\end{equation}
At $\delta K_{\rm loc}\simeq0.31$, this local protection is lost.
The quasiparticle weight then begins to decrease,
\begin{equation}
Z_{\rm QP}:
0.950
\rightarrow
0.936
\rightarrow
0.828
\quad
\text{for}
\quad
\delta K:
0.31
\rightarrow
0.41.
\end{equation}

At the C+1m level, this decrease represents transfer
of spectral weight from the coherent pole to the one-magnon
scattering continuum. Once higher-magnon sectors are included, the
same loss of local spiral stability allows repeated magnon
emission, producing the multi-magnon continuum observed in
Fig.~\ref{fig:polaron}(a) of the main text. The threshold
$\delta K_{\rm loc}$ is therefore the one-magnon precursor of the
magnon proliferation that destroys the coherent polaron.

The hierarchy
\begin{equation}
\delta K_E
=
0.170
<
\delta K_{\rm loc}
\simeq
0.31
\end{equation}
separates two physically distinct phenomena.
The first is a global variational crossing to a state with
$Z=0$, which is invisible in the single-hole spectral function.
The second is the loss of local stability of the spiral branch,
which coincides, within the numerical resolution, with the onset
of the reduction of $Z_{\rm QP}$. The critical momentum
$\mathbf K_c$ in Fig.~3(a) therefore marks the point at which the
spiral supporting the coherent quasiparticle becomes locally
unstable. The incoherent spectral weight beyond $\mathbf K_c$
arises from the ensuing proliferation of magnon excitations,
rather than from a transition to a distinct competing magnetic
order.

\end{document}